\documentclass[12pt]{article}
\usepackage{graphicx}
\date{}
\begin{document}

\title{CELLULAR AUTOMATION OF GALACTIC HABITABLE ZONE}

\author{B. VUKOTI\'C \lowercase{and} M. M. \'CIRKOVI\'C}



\maketitle
\begin{center}
\it
Astronomical Observatory, Volgina 7, 11160 Belgrade-74, Serbia\\\vspace{4mm}
E-mail: bvukotic@aob.rs, mcirkovic@aob.rs\\
\vspace{8mm}
\end{center}
\abstract{We present a preliminary results of our Galactic
Habitable Zone (GHZ) 2D probabilistic cellular automata models.
The relevant time-scales (emergence of life, it's diversification
and evolution influenced with the global risk function) are
modeled as the probability matrix elements and are chosen in
accordance with the Copernican principle to be well-represented
by the data inferred from the Earth's fossil record. With Fermi's
paradox as a main boundary condition the resulting histories of
astrobiological landscape are discussed.}

\section{INTRODUCTION}

Since its introduction about a decade ago (Gonzalez, Brownlee and
Ward 2001; cf.\ the review of Gonzalez 2005), the concept of
Galactic Habitable Zone (GHZ) has been gaining momentum as a
flexible platform for advanced astrobiological computer
simulations of the Galaxy. In a simplified version, GHZ is an
annular ring in the plane of the Galactic disk with sufficient
metallicity for rocky planets formation and satisfying other
habitability conditions such as dynamical stability and limited
number of catastrophic explosions (supernovae, gamma-ray bursts).
In the most restrictive approach so far (Lineweaver 2001) the
ring spans between 7 and 9 kpc, and is slowly widening while
moving outwards with cosmic time, mainly because of the Galactic
chemical evolution. In previous papers (e.g., \'Cirkovi\'c and
Vukoti\'c 2008), we have used simplified 1D numerical models in
order to derive most general applicability conditions for a
number of astrobiological problems, like Fermi's paradox or
Carter's anthropic argument. Here, we attempt to reach more
detailed picture, while relying on the same concept of GHZ to
investigate the present number of Galactic sites with high
astrobiological complexity (i.e., developed technological
civilization). Our work in this preliminary study is based on the
Probabilistic Cellular Automata (PCA) approach presented below.

In brief, the PCA (like any cellular automaton) is constructed as
a lattice of cells that evolve in discrete time steps (e.g.,
Ilachinski 2001). Each cell is allowed same several discrete
states. The state of the cell in the next time step depends on
the state of the cell itself and on the state of the surrounding
cells in the current time step. The evolution rules are defined
with the set of {\it transition probabilities}, thus defining the
stochastic subset of cellular automata. This is particularly
convenient for modelling the systems with large number of unknown
or poorly understood parameters such as GHZ. These parameters can
be easily implemented and changed via the transition
probabilities. Here, we will perform the PCA simulations of GHZ
with variable transition probabilities. The results and input
parameters will be considered in the frame of the Fermi's paradox
as a boundary condition.

Fermi's paradox -- the lack of traces or manifestations of
extraterrestrial civilizations despite the vast amount of time
since the formation of Galactic disk (for reviews see Webb 2001,
\'Cirkovi\'c 2009) -- can be viewed as a specific boundary
condition for the purpose of astrobiological modelling. Here we
consider two possibilities. In hard approximation the condition
assumes that there are no/few extraterrestrial societies or that
such civilizations are not found of expansion (see, \'Cirkovi\'c
2008) -- this can account for the lack of the contact so far. The
soft approximation puts less weight on a Fermi's paradox as a
boundary condition; Either we are living in the passive corner of
the Galaxy and are constantly being missed by extraterrestrial's
explorations (like an Amazon forest tribe, cf.\ Kinouchi 2001), or
our fellow Galactic cohabitants are keeping us unaware of their
presence (the zoo hypothesis). Each of the two presented
possibilities can be considered as a different set of transition
probabilities for the purpose of computer simulation input
parameters.

At present, we have no reasons to prefer one possibility over
another. The output of our simulations is given as the contour
plots of the developed civilization number at the end of the
simulation run in the transition probabilities phase space. In
the case of smooth contour plots there will still be no special
reason to favour one of the two scenarios given above. However,
if it turns out that contour plots show the fast changes over the
transition probability phase space it will mean that some values
of transition probabilities can give a plausible solution to the
'Great Silence' problem, in the light of the adopted model.

\section{MODEL DESCRIPTION}

The discrete cell states are: 0 - no life, 1 - simple life, 2 - complex life, 3 - technological civilization. The automation is performed on a rectangular grid with GHZ between $21^\mathrm{st}$ and $36^\mathrm{th}$ cell from the center of the grid, for the inner and outer radius of the annulus. One simulation step represents 5 Myr on the 0 - 10 Gyr time scale. The GHZ grid cells remain inactive until invoked to the state 0 according to the Earth-like planet formation rate probabilities distribution from Lineweaver (2001). In this simple model each cell is a representative of one stellar system with the main sequence host star life time chosen randomly for the $0.6 - 1.3~m_\odot$ mass range (the plausible range for habitable Solar-like systems). If the main sequence life time expires before the end of the simulation run the cell is made inactive until again invoked by the mentioned probability distribution before the simulation run ends.

Table 1 lists transition probabilities implemented in our model. First column gives the transition description and the last column gives the probability type. Evolution probabilities are driving the evolution of the cell residing stellar system towards higher atsrobiological complexity. Catastrophic probabilities are evolution impeding and are of complexity degrading nature, while colonization probabilities define the probability for the cell to be colonized by adjacent state 3 cells.

First, the probabilities of all possible transitions are calculated as:
\begin{equation}
p_{ij} = \frac{A_{ij}  t_i  t_\mathrm{res}}{t_{\mathrm{char}(ij)}},
\end{equation}
where $A_{ij}$ is the amplitude in the (0, 1) interval and is used to further control the $i\rightarrow j$ transition probability influence, $t_i$ is the time that cell have spent in the state $i$, $t_\mathrm{res}$ is the simulation time step and  $t_{\mathrm{char}(ij)}$ is the characteristic time of the $i\rightarrow j$ transition probability. At initialization each cell is assigned a set of random numbers from the (0, 1) interval with each element of the set corresponding to one transition. The set is reinitialized after each change of the cell state. If the transition probability is higher than a currently assigned corresponding random number then the transition is considered as pending. After this, all pending transition probabilities are stacked in an array. First probability is the first element of an array, while the sum of the second and first probability is the second element of the array and so on. The array is then normalized so the final element is one. Another random number is generated and compared with the elements of the normalized array. The transition that corresponds to the highest element of the normalized array that is smaller than generated random number, occurs.

\begin{table}
\begin{center}
\begin{tabular}{|c @{ $\rightarrow$}c|c c c|}
\hline
\multicolumn{2}{|c|}{Transition} & Char. time [Gyr] & Amplitude & Type \\
\hline
0 & 1 & 3 & 1.0 & evolution\\
1 & 2 & 0.6 & 1.0 & evolution\\
2 & 3 & 0.1 & 1.0 & evolution\\
3 & 0 & variable & variable & catastrophic \\
3 & 1 & variable & variable & catastrophic \\
3 & 2 & variable & variable & catastrophic \\
0 & 3 & variable & variable & colonization \\
1 & 3 & variable & variable & colonization \\
2 & 3 & variable & variable & colonization \\
\hline
\end{tabular}
\caption{Transition probabilities implemented in our simulations.}
\end{center}
\end{table}

\section{RESULTS}

\begin{figure}
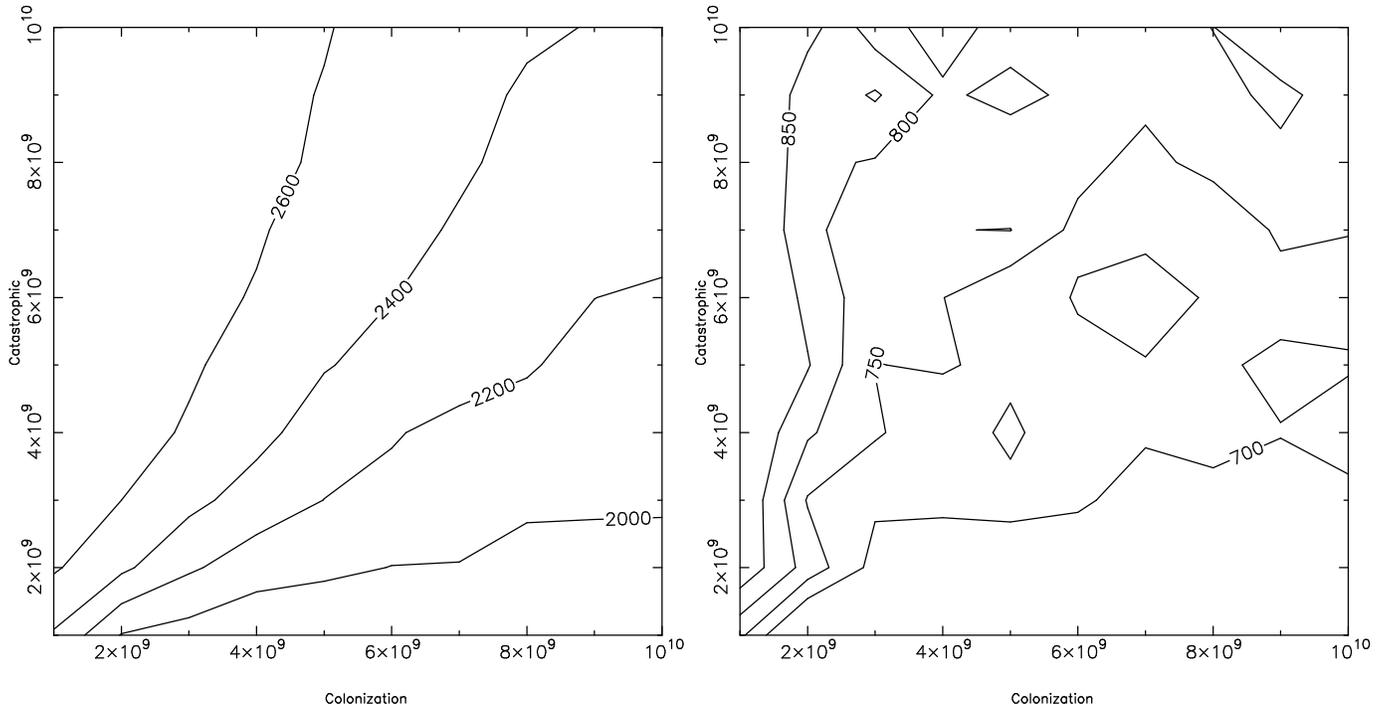

\includegraphics[width=0.52\textwidth, angle = -90]{smooth.eps}
\includegraphics[width=0.52\textwidth, angle = -90]{glitchy.eps}
\caption{The contour plots of the number of cells in state 3 at the end of the simulation run for various values of input transition probabilities characteristic times. The details are given in the text.}
\end{figure}

We have performed simulation runs for various amplitudes of Table 1 probabilities and characteristic times (ranging from $10^7 - 10^8$ yr with $10^7$ yr time-step, $10^8 - 10^9$ yr with $10^8$ yr time-step and $10^9 - 10^{10}$ yr with $10^9$ yr time-step). For the sake of brevity we present one of the smooth contour plots and one with the irregular contours (Figure 1). For the left hand side panel in Figure 1  transitions $3\rightarrow0$ and $3\rightarrow1$ were attenuated with $A_{30}=0.01$ and  $A_{31}=0.1$, respectively,  while the remaining transitions have amplitudes of 1.0, as is the case with the right hand side panel of Figure 1. Also, for the right hand side panel part of the Figure 1, $3\rightarrow1$ and $3\rightarrow2$ transitions characteristic times were kept fixed at $t_{\mathrm{char}(31)}=600$ Myr and $t_{\mathrm{char}(32)}=100$ Myr, respectively. The characteristic times of the remaining transitions were varied from $10^9$ to $10^{10}$ yr with the resolution of $10^9$ yr, for both panels. The values of $t_{\mathrm{char}(ij)}$ for colonization transitions are given on horizontal axis while catastrophic transitions values are on the vertical axis.

The irregular shape of the contours in the right hand side panel
implies that the method presented in this work can give plausible
solutions to the 'Great Silence' problem. Also, it demonstrates
the possible island-like nature of the "archipelago of
habitability" in the transition probabilities phase space. Given
the largely arbitrary character of the input parameters presented
here, and various model simplifications, no conclusions should be
made until the results of more comprehensive future studies;
obviously, more detailed results are contingent on improving
results from plethora of different disciplines, notably
paleobiology, (extrasolar) planetary sciences and computational
physics. Instead, this work should be viewed as a form of
justification for their undertaking and an illustration what
modern numerical methods, unfortunately largely neglected in this field of study so far, could achieve.\\

{\it \noindent Acknowledgements.} We thank Anders Sandberg,
Robert J. Bradbury, Branislav K. Nikoli\'c, Zoran Kne\v zevi\'c,
Nikola Bo\v zi\'c, and Damian Veal for kind encouragement,
technical help and useful discussions on a range of relevant
topics. This research has been supported by the Ministry of
Science of the Republic of Serbia through the project ON146012:
"Gaseous and stellar components of galaxies: interaction and
evolution".\\
\vspace{3mm}

{\noindent \bf References}\\

\'Cirkovi\'c, M. M. : 2008, {\it Journal of the British
Interplanetary Society}, {\bf 61}, 246-254.

\'Cirkovi\'c, M. M. : 2009,  {\it Serbian Astronomical
Journal}, {\bf 178}, 1.

\'Cirkovi\'c, M. M. and Vukoti\'c, B. : 2008, {\it Origin of
Life and Evolution of the Biosphere}, {\bf 38}, \indent 535.

Gonzalez, G. : 2005, {\it Origin of Life and Evolution of the
Biosphere}, {\bf 35}, 555-606.

Gonzalez, G., Brownlee, D., and Ward, P. : 2001, {\it Icarus}, {\bf 152}, 185.

Ilachinski, A. : 2001, {\it CELLULAR AUTOMATA: A Discrete Universe}, World Scientific \indent Publishing, Singapore.

Kinouchi, O. : 2001, ArXiv Condensed Matter e-prints, arXiv:cond-mat/0112137.

Lineweaver, C. H. : 2001, {\it Icarus}, {\bf 151}, 307.

Webb, S. : 2002, {\it Where is Everybody? Fifty Solutions to the
Fermi's Paradox\/}, Copernicus, \indent New York.


\end{document}